\documentclass[
aps,
nofootinbib,
prd,
superscriptaddress,
tightenlines,
notitlepage,
twocolumn,
showpacs,
floatfix
]{revtex4-2}
\usepackage{amsmath}
\usepackage{latexsym}
\usepackage{amsfonts}
\usepackage{graphicx}
\usepackage{mathrsfs}
\usepackage{epstopdf}
\usepackage{subfigure}
\usepackage[utf8]{inputenc}
\usepackage{CJK}
\usepackage{float}
\usepackage{color}
\usepackage{hyperref}
\usepackage{diagbox}
\usepackage{xcolor}
\usepackage{bm}
\usepackage{lipsum}
\hypersetup{
    colorlinks=true,
    linkcolor=red,
    citecolor=blue,
}

\begin{document}

\title{Unveiling the existence of nontensorial gravitational-wave
polarizations from individual supermassive black hole binaries with pulsar
timing arrays}

\author{Dicong Liang}
\email[Corresponding author: ]{dcliang@pku.edu.cn}
\affiliation{Department of Mathematics and Physics, School of Biomedical Engineering, Southern Medical University, Guangzhou 510515, China}
\affiliation{Kavli Institute for Astronomy and Astrophysics, Peking University,
Beijing 100871, China}

\author{Siyuan Chen}
\affiliation{Kavli Institute for Astronomy and Astrophysics, Peking University,
Beijing 100871, China}
\affiliation{Shanghai Astronomical Observatory, Chinese Academy of Sciences, Shanghai 200030, China}

\author{Chao Zhang}
\affiliation{School of Aeronautics and Astronautics, Shanghai Jiao Tong
University, Shanghai 200240, China}

\author{Lijing Shao}
\email[Corresponding author: ]{lshao@pku.edu.cn}
\affiliation{Kavli Institute for Astronomy and Astrophysics, Peking University,
Beijing 100871, China}
\affiliation{National Astronomical Observatories, Chinese Academy of Sciences,
Beijing 100012, China}


\begin{abstract}
With the strong evidence for a gravitational wave (GW) background in the
nanohertz frequency band from pulsar timing arrays, the detection of continuous
GWs from individual supermassive black hole binaries is already at the dawn.
Utilizing continuous GWs to test theories of gravity, especially to test the
polarizations of GWs is becoming more and more realistic.  In this theoretical study,
assuming a detection of signals from individual supermassive binary black holes,
we use the null stream to estimate the capability of identifying the
nontensorial polarizations of GWs.  We consider cases for the nontensorial
polarizations where the dipole radiation and quadrupole radiation dominate
separately.  With a frequentist method, we estimate the threshold of the
nontensor-to-tensor relative amplitude above which extra polarizations can
be detected.  We also conduct Bayesian analysis to estimate parameters with
the null stream data. Our treatment provides a data-analysis
methodology using the null stream to probe the nontensorial GW polarizations with pulsar timing arrays.
\end{abstract}

\maketitle

\section{Introduction}

Recently, evidence for a gravitational wave (GW) background was found by several
pulsar timing array (PTA) collaborations, which are the North American Nanohertz
Observatory for Gravitational Waves (NANOGrav) \cite{NANOGrav:2023gor}, the
European Pulsar Timing Array (EPTA) \cite{EPTA:2023fyk}, the Parkes Pulsar
Timing Array (PPTA) \cite{Reardon:2023gzh}, and the Chinese Pulsar Timing Array
(CPTA) \cite{Xu:2023wog}.  The detection of GWs by PTAs will enable us to probe the early Universe or
new physics encoded in nanohertz GWs \cite{Janssen:2014dka,
NANOGrav:2023hvm, EPTA:2023xxk, EuropeanPulsarTimingArray:2023egv, Li:2021qer, Zhang:2023lzt,
Yi:2023mbm, Yi:2023tdk, Wang:2023div, Wang:2023ost,
Chen:2024fir,Bi:2023ewq,Wu:2023rib,Liu:2023ymk,Liu:2023pau,Liu:2023hpw,Chen:2024twp}.

As an important property of GWs, the polarization content can be used to test
gravity theories.  There can be up to six polarizations in general metric
gravity theories \cite{Eardley:1973zuo,Eardley:1973br}. In addition to the
two tensor polarizations in general relativity, there can be scalar
polarizations and vector polarizations in modified gravity theories
\cite{Liang:2017ahj,Jacobson:2004ts,Liang:2022hxd}.  The response of a PTA to
the nontensorial polarizations and the corresponding overlap reduction function
in the GW background (GWB) search were discussed in a general framework in
Refs.~\cite{Chamberlin:2011ev, Boitier:2020rzg, Boitier:2020xfx, Hu:2022ujx,
Bernardo:2022rif}, and were studied in some specific theories
\cite{Hou:2017bqj,Gong:2018vbo,Gong:2018cgj,Liang:2021bct}.  Forecasts on
constraining extra polarizations with GWB signals were widely studied in
literature \cite{Gair:2015hra,Cornish:2017oic}.  Recent analysis on the real
dataset from NANOGrav suggests that the possibility of the existence of the
breathing mode cannot be excluded \cite{Chen:2023uiz,NANOGrav:2023ygs}.

Continuous GWs emitted from an individual supermassive binary black hole (SMBBH)
is another important target signal for PTAs \cite{Sesana:2008xk,Lee:2011et,
Janssen:2014dka, Rosado:2015epa,Kelley:2017vox,Becsy:2022zbu}, and GWs from the
loudest signals from individual SMBBHs are expected to be detected following the
detection of GWB \cite{Rosado:2015epa}.  Current data provide no evidence of
such a signal \cite{NANOGrav:2023pdq,EPTA:2023gyr}.  Nevertheless, there are a
few studies on constraining the extra polarizations of GWs from individual
SMBBHs \cite{Niu:2018oox,OBeirne:2019lwp}.

For the GW detection in the audio frequency band at $\sim 10^2$\,Hz, Bayesian model selection is generally used to do the polarization tests
\cite{Isi:2015cva, Isi:2017equ, Isi:2017fbj, LIGOScientific:2017ycc, LIGOScientific:2017ous, Callister:2017ocg,
LIGOScientific:2018czr,Takeda:2020tjj,Takeda:2021hgo}. 
Using Fisher matrix, gravitational-wave polarization tests with future ground-based detectors was studied in Ref.~\cite{Takeda:2019gwk}.
At the same time, the null stream method has been
developed to probe the existence of nontensorial polarizations for the
ground-based GW detector network and space-based GW detectors
\cite{Hagihara:2019ihn, Pang:2020pf, Wong:2021cmp, LIGOScientific:2020tif, 
LIGOScientific:2021sio, Hu:2023soi, Zhang:2021fha}.  Null streams can be
constructed by the special linear combination of the multiple data streams from
the detector network, such that the tensorial signals are completely eliminated
\cite{Guersel:1989th, Chatterji:2006nh, Chatziioannou:2012rf}.  Based on this
idea, there are also some studies on using null stream to localize the source in
the PTA community \cite{Zhu:2015tua,Zhu:2016zlx, Goldstein:2017qub,
Goldstein:2018rdr}.  In this work, for the first time we utilize the null
streams to estimate the detectability of nontensorial GWs with PTAs, which
provides a strategy in using the null stream to probe the nontensorial GW
polarizations for future real data analysis.
Throughout the paper, we only consider one GW source for each simulation.

The paper is organized as follows.  In Sec.~\ref{sec-pre}, we introduce some
basic concepts about extra polarizations, the waveform models, and the way to
construct null streams.  Next, we estimate the sensitivity of PTAs to the extra
polarization and obtain the detection threshold of the nontensor-to-tensor
relative amplitude with a frequentist method in Sec.~\ref{sec-fre}.  Then we use
the null stream to estimate the parameters with Bayesian inference in
Sec.~\ref{sec-bay}.  Final discussions are presented in Sec.~\ref{sec-dis}.

\section{Methodology}
\label{sec-pre}

In this section we provide the settings used in this work, including the extra
GW polarizations in PTAs in Sec.~\ref{sec:pol}, GW waveforms in
Sec.~\ref{sec:waveform}, and the null stream in Sec.~\ref{sec:ns}.

\subsection{Extra polarizations}\label{sec:pol}

We first introduce the orthonormal coordinate system
\begin{align}
    \hat{\mathbf{\Omega}}=( & \sin\theta \cos\phi,\sin\theta\sin\phi,\cos\theta) \, ,
    \nonumber \\
    \hat{\mathbf{m}}= ( &\cos\theta\cos\phi\cos\psi-\sin\phi\sin\psi, 
    \nonumber \\
    & \cos\theta\sin\phi\cos\psi+\cos\phi\sin\psi ,-\sin\theta\cos\psi) \, ,
    \nonumber \\
    \hat{\mathbf{n}}= (&-\cos\theta\cos\phi\sin\psi-\sin\phi\cos\psi, \nonumber \\
    &-\cos\theta\sin\phi\sin\psi +\cos\phi\cos\psi,
    \sin\theta\sin\psi) \, ,
\end{align}
where $\theta$ and $\phi$ denote the propagation direction of the GWs (i.e.
$-\hat{\mathbf{\Omega}}$ points towards the GW source), and $\psi$ is the
polarization angle.  Then the basic tensors for the six polarizations (denoted
as `$+$', `$\times$', `$x$', `$y$', `$b$', `$l$') are
\begin{align}
   & e^+_{ij} = \hat{m}_i \hat{m}_j - \hat{n}_i \hat{n}_j \, , 
   \quad
    e^\times_{ij} = \hat{m}_i \hat{n}_j + \hat{n}_i \hat{m}_j\, ,
    \nonumber \\
   & e^x_{ij} = \hat{m}_i \hat{\Omega}_j + \hat{\Omega}_i \hat{m}_j \, ,
    \quad
    e^y_{ij} = \hat{n}_i \hat{\Omega}_j  + \hat{\Omega}_i \hat{n}_j\, ,
    \nonumber \\
   & e^b_{ij} = \hat{m}_i \hat{m}_j + \hat{n}_i \hat{n}_j \, ,
    \quad
    e^l_{ij} = \hat{\Omega}_i \hat{\Omega}_j \, .
\end{align}

We only consider the case where all the possible polarizations propagate at the
speed of light and are monochromatic.  Then, the timing residuals $r_A$ of a
pulsar induced by polarization $A$ are given by
\begin{align}
r_A =F_A h_A,     
\end{align}
with the strain $h_A$ and the response function $F_A$
\cite{Romano:2016dpx,Romano:2023zhb}
\begin{align}
    F_A  =\frac{1}{4\pi i f_A} \frac{\hat{p}^i \hat{p}^j e^A_{ij}
    (\hat{\mathbf{\Omega}}) }{1+\hat{\mathbf{\Omega}}\cdot \hat{\mathbf{p}}} 
    \left[ 1-e^{-i 2\pi f_A L_p(1+ \hat{\mathbf{\Omega}}\cdot \hat{\mathbf{p}})
    /c } \right] .
\label{eq-response}
\end{align}
Here $\hat{\bf p}$ is the unit vector pointing in the direction to the pulsar,
$L_p$ is the distance between the pulsar and the Earth, $f_A$ is the frequency
of the GW polarization mode $A$.  Even for a circular binary, the extra
polarizations can have different frequency evolution from the tensor
polarizations \cite{Chatziioannou:2012rf}.  We also define the combined response
to the tensor and vector modes as $|F_t|^2 \equiv |F_+|^2 + |F_\times|^2$, and
$|F_v|^2 \equiv |F_x|^2 + |F_y|^2$. 

In this work, we consider $N_p=49$ pulsars from the International Pulsar Timing
Array data release 1 (IPTA-DR1) \cite{Verbiest:2016vem} and the distances of
these pulsars are inquired from the Australia Telescope National Facility (ATNF)
pulsar catalogue\footnote{
\url{https://www.atnf.csiro.au/people/pulsar/psrcat/}} \cite{Manchester:2004bp}.
As we can infer from the ATNF catalogue, the sky position of pulsars, $\hat{\bf p}$, can be measured to a high precision while the distances of the pulsars, $L_p$, cannot be measured very accurately so far. In this theoretical study, we want to estimate the capability of identifying the extra modes after the detection of the signals. Thus, we further assume theoretical knowledge of the frequency, $f$, and the source location, $\hat{\mathbf{\Omega}}$, so that we have precise knowledge of the response function. 

\begin{figure*}
 \includegraphics[width=\linewidth]{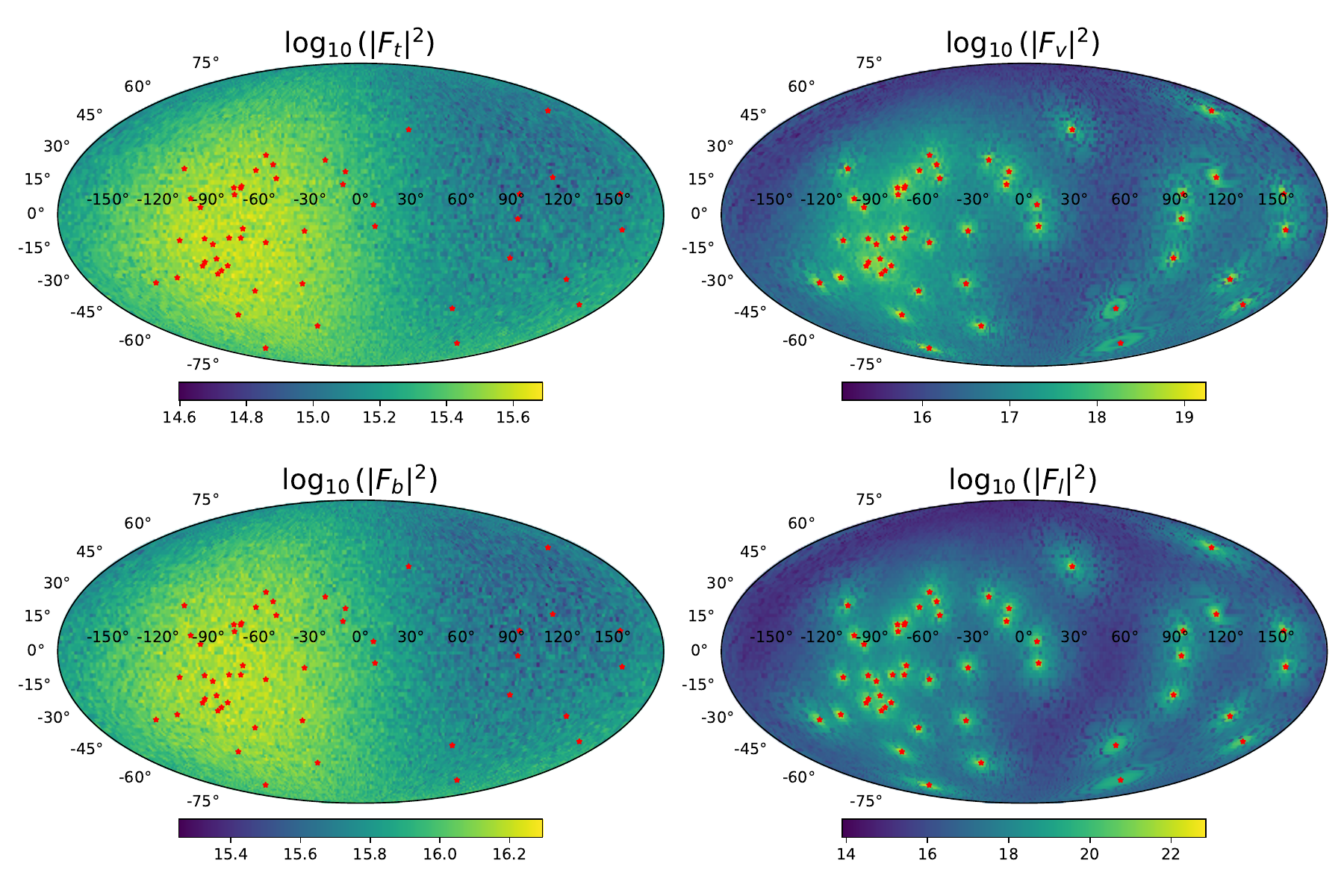}
 \caption{The overall response of the 49 pulsars (denoted with red stars) at
 different sky locations.  Here, we set the frequency of the tensor modes as $20
 \, \text{nHz}$ while the frequency of all the other modes is $10 \,
 \text{nHz}$.  }
\label{fig-response}
\end{figure*}

For now we only consider white Gaussian noise in the
data and we further assume the timing residuals of all the pulsars have the same
variance $\sigma_t= 100 \, \rm{ns}$ for simplicity.  The overall angular
response of these 49 pulsars to different polarizations, i.e., $ |F_A|^2 =
\sum_{i=1}^{N_p} |F_{i,A}|^2 $, is shown in Fig.~\ref{fig-response}.  The 49
pulsars are denoted with red stars in the figure.  Here, we assume that the
frequency of the extra polarizations is half of that of tensor polarizations,
corresponding to the dipole radiation.  To consider different timing residual
variances we can get the response by a re-weighted summation of the response of
each pulsar.

The most sensitive regions for the transverse modes (i.e., the tensor modes and
the breathing mode) are the left hemisphere, where most of the pulsars are
located.  While the most sensitive region for the vector modes and the
longitudinal mode are the ``islands" surrounding each pulsar.
This is because
that the response gets boosted for these modes when the angular separation
between the GW source and the pulsar is small \cite{OBeirne:2019lwp}.  Another
important feature of the sensitivity skymap is that, the response oscillates
quickly over the sky location due to the ``pulsar term'', which is the second
term in the square bracket in Eq.~\eqref{eq-response}.

\subsection{Waveform model}\label{sec:waveform}

For the two tensor modes, we parameterize the waveform as follows 
\begin{align}
    h_+=&  h_0 \frac{1+\cos^2\iota}{2} \exp \big[i 
 (2\pi f_0 t + 2 \Phi_0 ) \big] ,
    \nonumber \\
    h_\times=& -i h_0 \cos\iota \exp \big[i(2\pi f_0 t + 
 2 \Phi_0 )\big] ,
\end{align}
where $\iota$ is the inclination angle, $\Phi_0$ is the initial orbital phase of the binary, $h_0$ is the fiducial
amplitude, which depends on the binary masses, frequency and luminosity
distance.  Throughout the paper, we adopt the frequency $f_0 = 20 \, \text{nHz}$ and fix $h_0=
10^{-14}$, which is a bit smaller than the current upper limit estimated by
NANOGrav \cite{NANOGrav:2023pdq} and EPTA \cite{EPTA:2023gyr}.

For the extra polarizations, we consider two cases.  In the first case (Case 1),
we assume the dipole emission dominates as in Ref.~\cite{OBeirne:2019lwp}. The
parameterized waveforms are given by
\begin{align}
    h_x = & A_v h_0 \cos\iota \exp[i(\pi f_0 t + \Phi_0 )] ,   
    \nonumber \\
    h_y = & -i A_v h_0 \exp[i(\pi f_0 t + \Phi_0 )] ,  
    \nonumber \\
    h_b = & A_b h_0 \sin \iota \exp[i(\pi f_0 t + \Phi_0 )] , 
    \nonumber \\
    h_l = & A_l h_0 \sin\iota \exp[i(\pi f_0 t + \Phi_0 )] .
    \label{eq-waveform1}
\end{align}
Here, $A_v$, $A_b$, and $A_l$ are the relative amplitudes for the vector mode,
breathing mode and longitudinal mode respectively.  Notice that, in this case,
the frequency of the extra modes is only half of that of the tensor modes. 

For the second case (Case 2), we assume that quadrupole radiation dominates.
Then we follow \citet{Takeda:2018uai} to parameterize them as
\begin{align}
    h_x=& -A_v h_0\frac{\sin2\iota}{2} \exp \big[i(2\pi f_0 t + 2\Phi_0 ) \big] ,
    \nonumber \\
    h_y=& i A_v h_0 \sin\iota  \exp \big[i(2\pi f_0 t + 2\Phi_0 ) \big] ,
    \nonumber \\
    h_b=&- A_b h_0 \frac{\sin^2\iota}{2} \exp \big[i(2\pi f_0 t + 2\Phi_0 ) \big] ,
    \nonumber \\
    h_l=& A_l h_ 0 \frac{\sin^2\iota}{2} \exp \big[i(2\pi f_0 t + 2\Phi_0 ) \big] .
    \label{eq-waveform2}
\end{align}
In this case, all the extra polarizations have the same frequency as the tensor
modes.  Considering that $f_A$ is in the denominator of the response function
[see Eq.~\eqref{eq-response}], the response to the extra polarizations in Case 2
is weaker than that in Case 1.  We can also notice that the dependence on the
inclination angle is different between these two cases.

Following \citet{Goldstein:2017qub}, we assume $N_{\text{TOA}}=300$ times of
arrival (TOAs),  evenly distributed in time with a cadence of $\Delta t= 10^6 \,
\text{s}$. Thus, the total observation time is around 9.5 years.

\subsection{Null stream} \label{sec:ns}

Now we introduce how to construct a null stream \cite{Zhu:2015tua}.  In
general relativity, we can write the $N_p$ data streams in the matrix as follows
\begin{align}
    \left(  \begin{array}{cccc} 
       \mathbf{d}_1    \\
       \mathbf{d}_2   \\
       \vdots   \\
       \mathbf{d}_{N_p}    \\ 
    \end{array}  \right) = \left(  \begin{array}{cccc} 
       F_1^+  & F_1^\times  \\
       F_2^+  & F_2^\times  \\
       \vdots & \vdots  \\
       F_{N_p}^+  & F_{N_p}^\times  \\ 
    \end{array}  \right) 
    \left(  \begin{array}{cc} 
       \mathbf{h}^+    \\
       \mathbf{h}^\times  \\
    \end{array}  \right) +
    \left(  \begin{array}{cccc} 
       \mathbf{n}_1    \\
       \mathbf{n}_2   \\
       \vdots   \\
       \mathbf{n}_{N_p}    \\ 
    \end{array}  \right) ,
\end{align}
where the bold symbols indicate time series. We can also write it in a more compact form
\begin{align}
    \mathbf{d} = \mathbf{r}  +\mathbf{n} = \mathbf{F}_t \mathbf{h}_t +\mathbf{n} ,
\end{align}
where $\mathbf{F}_t = (\mathbf{F}^{+} , \ \mathbf{F}^{\times} )$.  We perform
the singular value decomposition for $\mathbf{F}_t$ as in
Ref.~\cite{Zhu:2015tua} to get 
\begin{align}
    \mathbf{F}_t  = \mathbf{U}\mathbf{S}\mathbf{V}^\dagger , \quad 
    \mathbf{S}= \left( \begin{array}{ccccc}
        s_1 & 0  \\
        0  & s_2 \\
        0  &  0 \\
        \vdots & \vdots \\
        0 & 0
    \end{array} \right),
\end{align}
where $\mathbf{U}$ and $\mathbf{V}$ are unitary matrices, $s_1$ and $s_2$ are
the singular values and the symbol $\dagger$ denotes conjugate transpose.
Projecting the matrix $\mathbf{U}^\dagger$ onto the original data streams, we finally obtain the new data streams
\begin{align}
    \Tilde{\mathbf{d}} \equiv \mathbf{U}^\dagger \mathbf{d} 
    =  \left(\begin{array}{ccccc}
         s_1 (\mathbf{V}^\dagger \mathbf{h}_t)_1 +\tilde{\mathbf{n}}_1 \\
         s_2 (\mathbf{V}^\dagger \mathbf{h}_t)_2 +\tilde{\mathbf{n}}_2 \\
         \tilde{\mathbf{n}}_3 \\
         \vdots \\
         \tilde{\mathbf{n}}_N
    \end{array} \right), 
\end{align}
where $\tilde{\mathbf{n}} = \mathbf{U}^\dagger \mathbf{n}$.  Now, the new data
streams, $\tilde{\mathbf{d}}_k$ with $k\geq 3$, are the so-called null streams since they
do not contain any tensorial signals.  Notice that $\mathbf{U}$ is a unitary
matrix, which does not change the statistical properties of the noise \cite{Zhu:2015tua,Zhu:2016zlx}.  In this
work, we consider the statistic $X$ related to the null streams, which is
defined as
\begin{align}
    X =  \sum_{k=3}^{N_p} \sum_n^{N_{\rm TOA}} \frac{1}{\sigma_t^2}
    \big| \tilde{d}_k(t_n) \big|^2.
\end{align}
Considering white noise only, the statistic $X$ follows a $\chi^2$ distribution
with degrees of freedom $N_{\rm dof}=(N_p-2) \times N_{\rm TOA}$.

Notice that the construction of these null streams requires precise knowledge of the response function matrix $\mathbf{F}_t$.
Achieving this in practice may be challenging due
to uncertainties in the quantities used to compute it.
Thus, it is an idealized assumption throughout the paper and we hope to relax it in the future study.

\section{Frequentist analysis}
\label{sec-fre}


If there are extra polarizations, then the data streams can now be written as
\begin{align}
    \mathbf{d} = \mathbf{F}_t \mathbf{h}_t + \mathbf{F}_e \mathbf{h}_e  +\mathbf{n} .
\end{align}
Here, the subscript $e$ denotes any extra polarization. After the projection by
$\mathbf{U}^\dagger$, the new tensor-polarization-null streams $\mathbf{d}_k \
(k\geq 3) $ still contain the component $( \mathbf{U}^\dagger \mathbf{F}_e
\mathbf{h}_e  )_j$, i.e.,
\begin{align}
    \Tilde{\mathbf{d}} 
    =  \left(\begin{array}{ccccc}
         s_1 (\mathbf{V}^\dagger \mathbf{h}_t)_1 +\tilde{\mathbf{n}}_1 
         + ( \mathbf{U}^\dagger \mathbf{F}_e \mathbf{h}_e  )_1  \\
         s_2 (\mathbf{V}^\dagger \mathbf{h}_t)_2 +\tilde{\mathbf{n}}_2 
         + ( \mathbf{U}^\dagger \mathbf{F}_e \mathbf{h}_e  )_2 \\
         \tilde{\mathbf{n}}_3 
         +  ( \mathbf{U}^\dagger \mathbf{F}_e \mathbf{h}_e  )_3 \\
         \vdots \\
         \tilde{\mathbf{n}}_N
         + ( \mathbf{U}^\dagger \mathbf{F}_e \mathbf{h}_e  )_{N_p}
    \end{array} \right) .
\end{align}
In this case, $X$ no longer follows the $\chi^2$ distribution and the deviation
can be significant if the extra polarization is strong enough.  We assign a
$p$-value to quantify its deviation from the $\chi^2$ distribution, defined as
\begin{align}
    p= \int_X^\infty \chi^2_{N_{\rm dof}}(x)dx.
\end{align}

As we can see in Fig.~\ref{fig-response}, the angular response is quickly
oscillating over the sky location due to the existence of the pulsar term. 
Thus, for the extra polarizations with the same amplitude, the effects on the
TOA can vary significantly for GW sources at different sky locations.  In other
words, $X$ can vary significantly for different sky locations.  To check the
average sensitivity of a PTA to the extra polarizations, we do some simulations
by injecting the extra polarizations separately.
Notice that, in all the following simulations, we injected both tensorial polarizations and the extra polarizations in the data.

Specifically, we first draw $\cos \iota$ from uniform distribution $U[0,1]$ and
draw $\psi$ and $\Phi_0$ from uniform distribution $U[0,2\pi)$.  After fixing
$\cos\iota$, $\psi$ and $\Phi_0$, we simulate $10^4$ uniformly spatial
distributed sources, i.e., $\cos\theta$ follows uniform distribution $U[-1,1]$
and $\phi$ follows uniform distribution $U[-\pi,\pi)$.  Then, we calculate the
$p$-value for these $10^4$ random sources separately with the same relative
amplitude $A$.  Finally, we calculate the fraction $\mathcal{F}$ of the events whose
$p<1.35\times 10^{-3}$, i.e.
\begin{align}
    \mathcal{F}=\frac{ N(p<1.35\times 10^{-3})}{ 10^4 },
\end{align}
We get the fraction $\mathcal{F}$ for a given $A$ for different polarizations in each
simulation.  It is an estimation of the size of the fraction of the sky
locations where we can detect extra polarizations with $>3\sigma$ significance
for the specific relative strength.  It can also be interpreted as the
probability of detecting nontensorial polarizations for a source randomly
distributed on the sky.

\begin{figure*}
 \includegraphics[width=\linewidth]{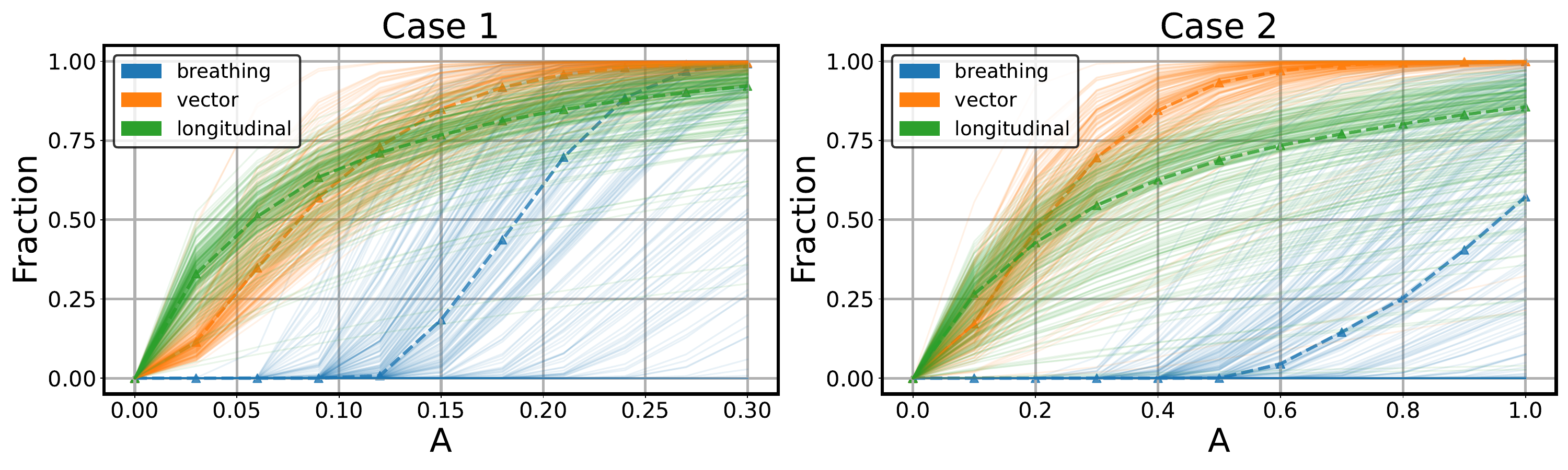}
 \caption{The fraction of sky locations we can detect extra polarizations with
 $>3\sigma$ confidence level at a given relative amplitude calculated for 200
 simulations.  Each thin line represents one simulation and the dotted line
 shows the median over the 200 simulations. }
\label{fig-location}
\end{figure*}

We perform 200 simulations and the results are denoted with thin lines  in
Fig.~\ref{fig-location} where different colors represent different injected
polarizations.  For the 10 choices of $A$, the median value of $\mathcal{F}$ of the 200
simulations are denoted as triangles in the figure.  The variance is due to the
different choices of $\iota$, $\psi$ and $\Phi_0$.  Notice that we choose
smaller amplitudes for simulations in Case 1 than that in Case 2, since the
response is stronger in Case 1 than that in Case 2.  In addition, the factor
corresponding to the inclination in the waveform is larger in Case 1 than that in Case 2.

Overall, the fraction decreases when the relative amplitude gets smaller and the
decline of the longitudinal mode is the slowest.  The variance of the breathing
mode is quite large here, since its strength strongly depends on the
inclination.  As shown in Fig.~\ref{fig-response}, the response at the brightest
regions is so strong that it dominates the contribution to the strength. Thus,
the effects of inclination are weaker for the longitudinal mode, although it has
the same inclination dependence as the breathing mode.

\begin{figure*}
\includegraphics[width=\linewidth]{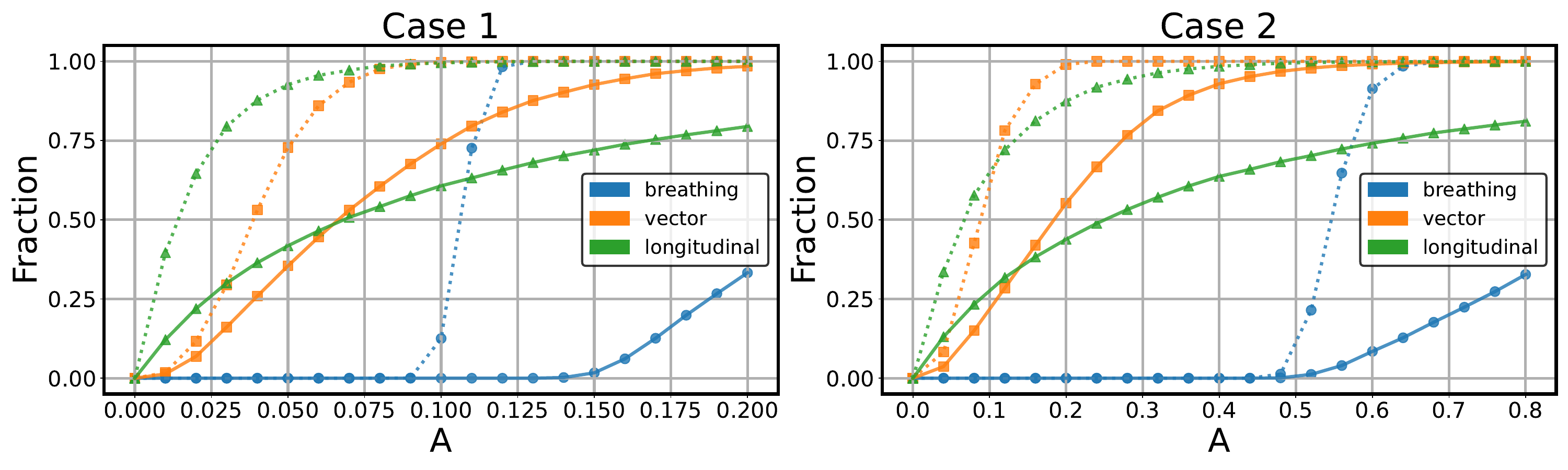}
\caption{ Similar to Fig.~\ref{fig-location}. The results of 49 pulsars and 200
pulsars are denoted with solid lines and dashed lines respectively. }
\label{fig-morepulsars}
\end{figure*}

Now, we extend our analysis to the cases with more pulsars.  
The Square Kilometre Array (SKA) will be built in the southern hemisphere of the Earth, adding a much larger number of observable pulsars \cite{Carilli:2004nx}. Thus, we expect a more isotropic distribution of pulsars in the future.
Here, we follow
\citet{Speri:2022tua} to create an array of 200 pulsars with Galaxy distribution
on the sky.  The distances of these pulsars follow a Gaussian distribution,
whose mean and variance are the same as those of the 49 pulsars in IPTA-DR1.  We
choose 20 sets of relative amplitude uniformly and set $\iota$, $\psi$ and
$\Phi_0$ to be $\pi/3$ for all the GW sources.  We again simulate $10^4$ sources
with different sky locations and calculate the fraction.  The results are shown
in Fig.~\ref{fig-morepulsars}.  The results of the 49 pulsars and 200 pulsars
are denoted with solid lines and dotted lines respectively.  Generally speaking,
when $A>0.125$ in Case 1 or $A>0.65$ in Case 2, the breathing mode can be
detected at $3\sigma$ level for almost all sky locations if there are 200
pulsars.  For the vector and longitudinal modes, more pulsars mean more
sensitive ``islands'', leading to significant improvement in the sky coverage. 

In the above analysis, we perform the singular value decomposition for the matrix
$\mathbf{F}_t= (\mathbf{F}^{+}, ~ \mathbf{F}^{\times} ) =
\mathbf{U}\mathbf{S}\mathbf{V}^\dagger$ to project out the tensor modes only.
Thus, we can only know that there exist nontensorial polarizations if we find
a significant deviation between the null stream and noise, but we cannot tell
which kind of extra polarizations are present.  However, we can construct $
\hat{\mathbf{U}}_b $ such that $(\mathbf{F}^{+} , ~ \mathbf{F}^{\times}  , ~
\mathbf{F}^{x}  , ~ \mathbf{F}^{y}  , ~ \mathbf{F}^{l} ) = \hat{\mathbf{U}}_b
\mathbf{S}_b \mathbf{V}^\dagger_b $.  Then we can get the new data streams
$\tilde{ \mathbf{d} }_b = \hat{\mathbf{U}}^\dagger_b  \mathbf{d}$, where 44 of
them contain only the breathing mode and the noise.  Using the same method, we
can obtain 45 data streams with only vector modes and 44 data streams with only
longitudinal modes, in addition to the noise. 

With these new null streams, we can define the injected SNR
\cite{OBeirne:2019lwp} as
\begin{align}
    \rho_{\rm inj} = \sum_{k=\ell}^{N_p} \sum_n^{N_{\rm TOA}}
    \frac{1}{\sigma_t^2} \big| ( \mathbf{U}^\dagger \mathbf{F}_e
\mathbf{h}_e  )_k(t_n) \big|^2,
\end{align}
where $\ell=4$ for the breathing and longitudinal modes, and $\ell=5$ for the
vector modes.  If the SNR is large enough, we expect large deviation of the null
stream from the noise and a corresponding extremely small $p$-value. In such a case we can claim strong evidence for the existence of the specific extra
polarizations.

\begin{figure*}
  \includegraphics[width=0.8\textwidth]{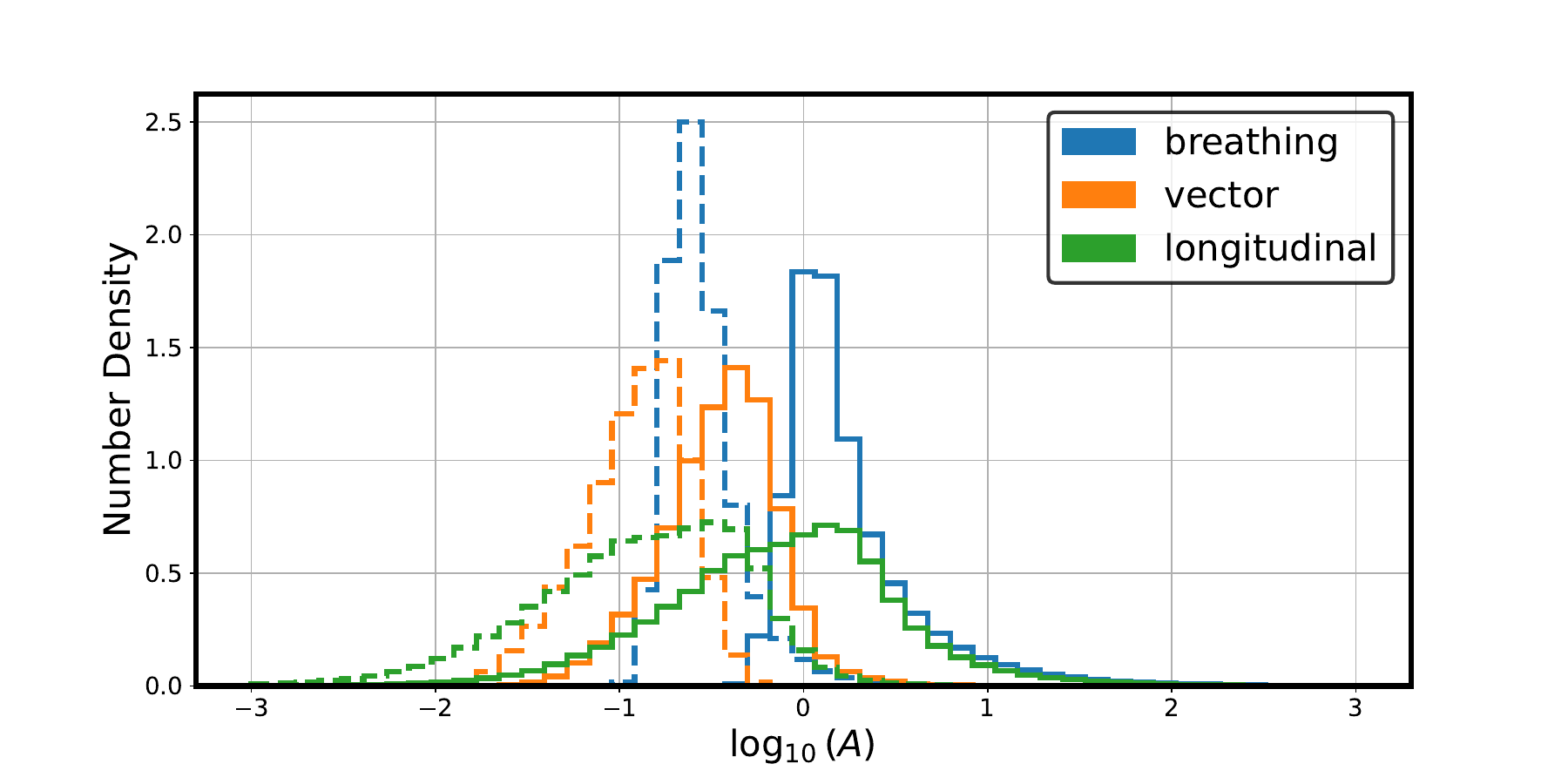}
 \caption{Distribution of the logarithm of relative amplitude for specific extra
 modes to produce an injected SNR $\rho_{\rm inj}=22$ after projecting out all
 the other modes in $2\times 10^6$ simulations. The results for Case 1 and Case
 2 are denoted with dashed and solid lines respectively. }
\label{fig-amplitude}
\end{figure*}

Now, we want to estimate how large the relative amplitude should be to detect
the specific extra polarizations.  First, we generate a random source, whose
angles, i.e. $\theta$, $\phi$, $\iota$, and $\Phi_0$ have the same distribution
as in the previous analysis.  Then, we find the threshold value of $A$ that
corresponds to an injected SNR $\rho_{\rm inj}= 22$, which is approximately
equivalent to a $3\sigma$ confidence level.  We generate $2\times 10^6$ random
sources, and the histogram of the threshold $A$ of these $2\times 10^6$ simulations
are shown in Fig.~\ref{fig-amplitude}. The dashed and solid lines represent the
results for Case 1 and Case 2 respectively.

For randomly distributed sources, the median values for the relative amplitude
to produce a signal with $\rho_{\rm inj}= 22$ in the null streams are $A_b=
0.26$, $A_v=0.13$, $A_l=0.15$ for Case 1, and $A_b=1.36$, $A_v=0.37$, $A_l=0.81$
for Case 2.  As explained previously, the expected amplitude is smaller in Case
1 than that in Case 2, due to its inclination dependence in the waveform and
frequency dependence in the angular response function.  It should be noted that
the variance of the relative amplitude of the longitudinal mode is very large. 
This is not surprising, as we can see in Fig.~\ref{fig-response}, that there can
be up to seven orders of magnitude difference in the response function for the
longitudinal mode.

\section{Bayesian analysis}
\label{sec-bay}

In the previous section, we estimated the average sensitivity of PTAs to the
extra polarizations and quantified the  relative amplitude threshold for
detection.  Now, we study how well we can recover the parameters of signals of
extra polarizations with the null streams.  According to Bayes' theorem, given
data $\mathbf{D}$, the posterior distribution of parameters
$\boldsymbol{\theta}$ is proportional to the product of the prior distribution
and the likelihood,
\begin{align}
    P(\boldsymbol{\theta}| \boldsymbol{D} ) \propto  P(\boldsymbol{\theta})
    P(\boldsymbol{D}| \boldsymbol{\theta} ) .
\end{align}
Here, the parameters we recover are
\begin{align}
    \boldsymbol{\theta} =\{ A_b, \ A_v, \ A_l, \ \iota, \ \psi, \ \Phi_0  \}.
\end{align}
As for the prior distribution, we assume $\log_{10} A_b$, $\log_{10} A_v$,
$\log_{10} A_l$ follow $U[-10,3]$, $\psi$ and $\Phi_0$ follow $U[0,2\pi)$ and
$\cos\iota$ follows $U[0,1]$.  After subtracting the waveform of extra polarizations $\mathbf{h}_e(\boldsymbol{\theta})$,  the residual data $
\mathbf{d}_s$,
\begin{align}
    \mathbf{d}_s = \mathbf{d} - \mathbf{F}_e \mathbf{h}_e ,
\end{align}
contain tensor modes and noise.  Then, the corresponding streams after
projection of $\mathbf{U}^\dagger$ are 
\begin{align}
    \tilde{ \mathbf{d} }_{s}= \mathbf{U}^\dagger \mathbf{d}_s .
\end{align}
Now, under the assumption of Gaussian noise, the log-likelihood with null
streams can be written as
\begin{align}
    \mathcal{L} \propto -\frac{1}{2} \sum_{k=3}^{N_p} \sum_n^{N_{\rm TOA}}
    \frac{1}{\sigma_t^2} \big|\tilde{d}_{s,k}(t_n) \big|^2 .
\end{align}

For all the simulations here, we fixed the source location to be $\theta=\pi/3$
and $\phi=-\pi/3$, and set $\iota=3\pi/5$, $\psi=4\pi/3$, $\Phi_0=1.2\pi$ for
both Case 1 and Case 2.  We inject the vector modes, breathing mode and
longitudinal mode separately, assuming that the other modes are zero.  For Case
1, the injection has $A_b= 1.2\times 10^{-1}$, $A_v=1.3\times 10^{-2}$ and $A_l=
2.9\times 10^{-4}$; for Case 2, the injection has $A_b= 6.6\times 10^{-1}$,
$A_v=1.9\times 10^{-2}$ and $A_l= 1.1\times 10^{-3}$.  For all these injections,
the relative amplitudes are chosen such that the corresponding injected SNR in
the null streams is approximately 22, as in the previous section. 

\begin{figure*}
 \includegraphics[width=\linewidth]{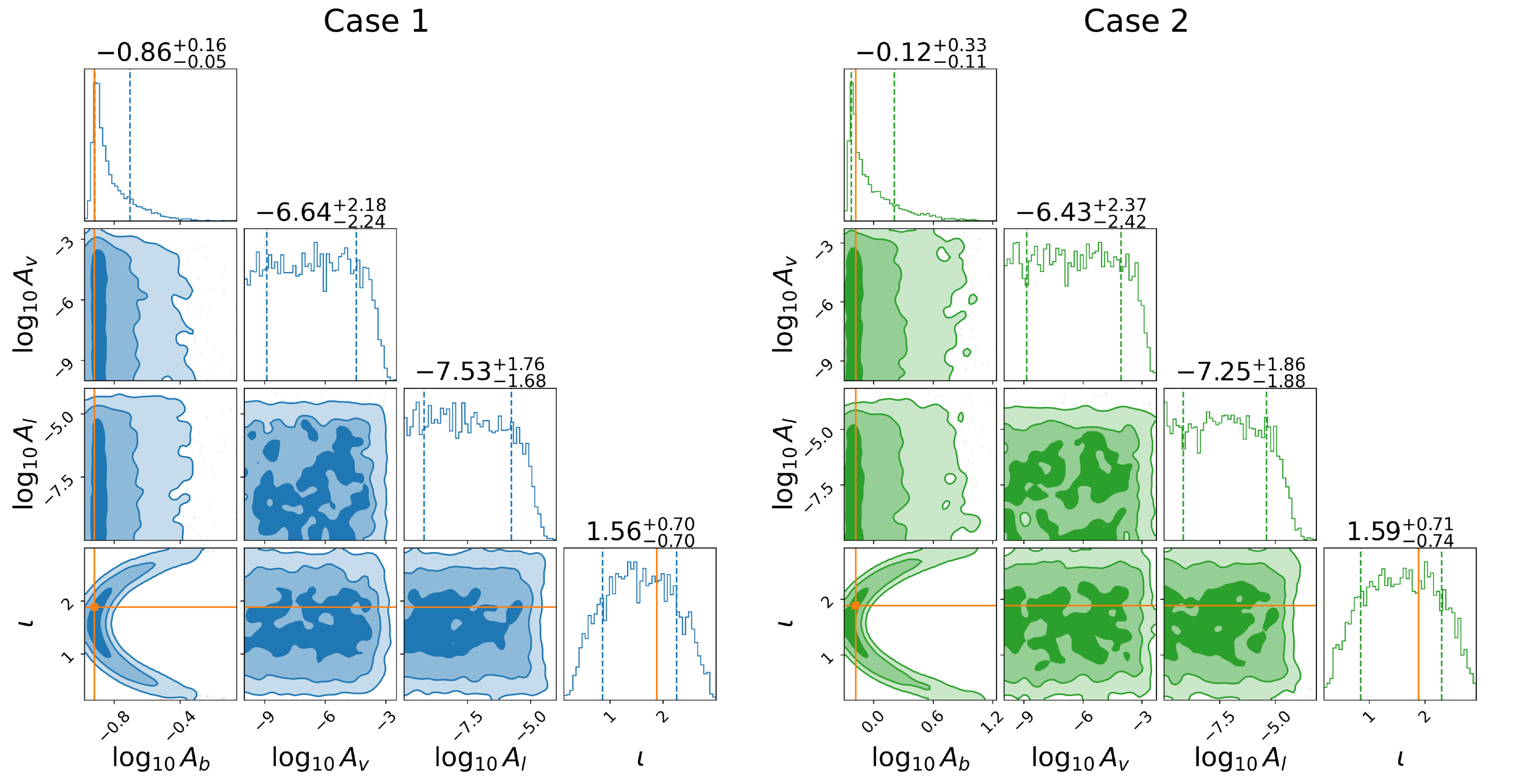}
 \caption{ Parameter estimation results with Bayesian inference (left panel for
 Case 1 and right panel for Case 2). Here, we inject
 breathing mode only. The orange dots/lines in the
 corner/histogram plots show the injected values. }
\label{fig-bayes-b}
\end{figure*}

\begin{figure*}
 \includegraphics[width=\linewidth]{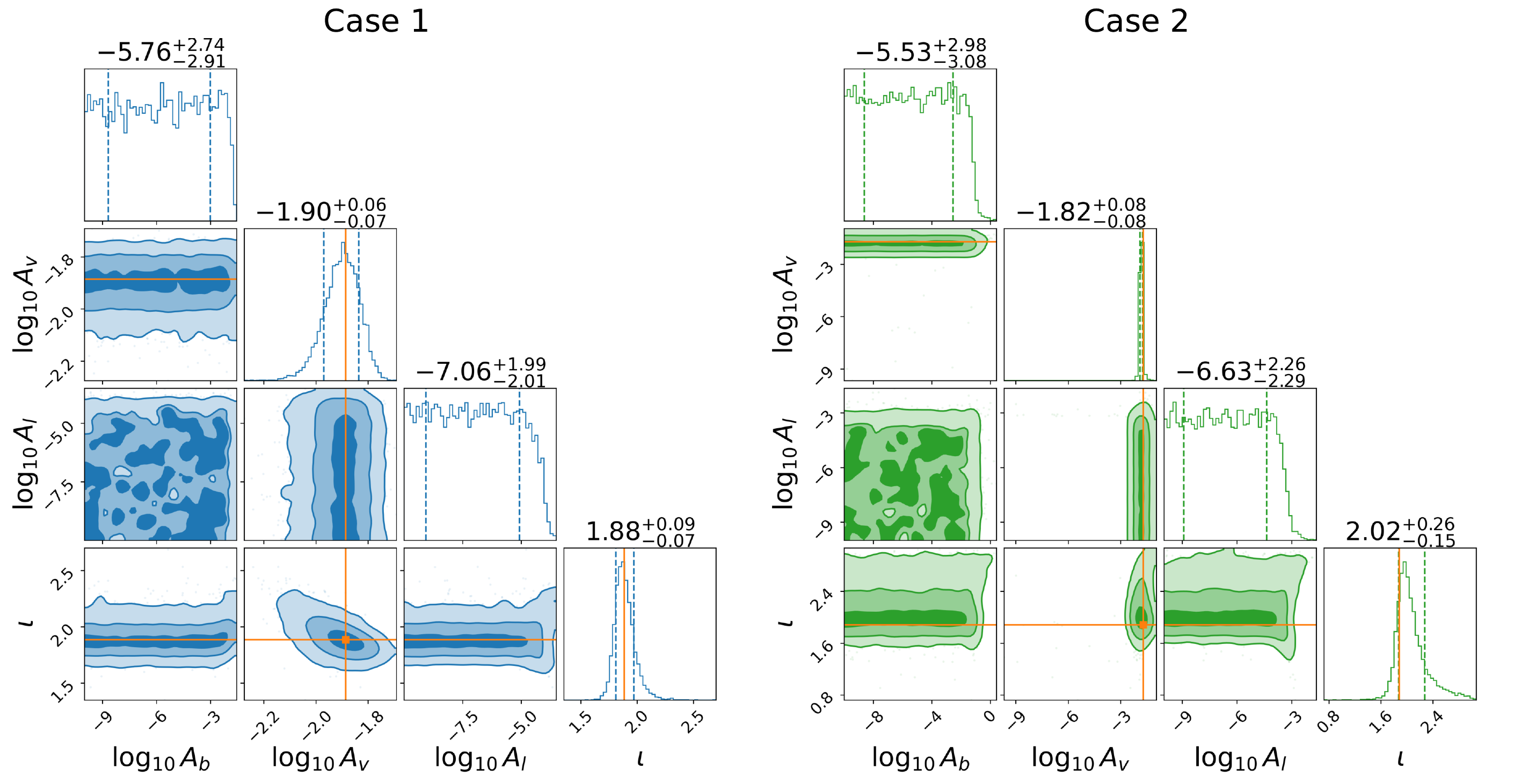}
 \caption{  Similar to Fig.~\ref{fig-bayes-b}. Here, we inject vector modes only.   }
\label{fig-bayes-v}
\end{figure*}

\begin{figure*}
 \includegraphics[width=\linewidth]{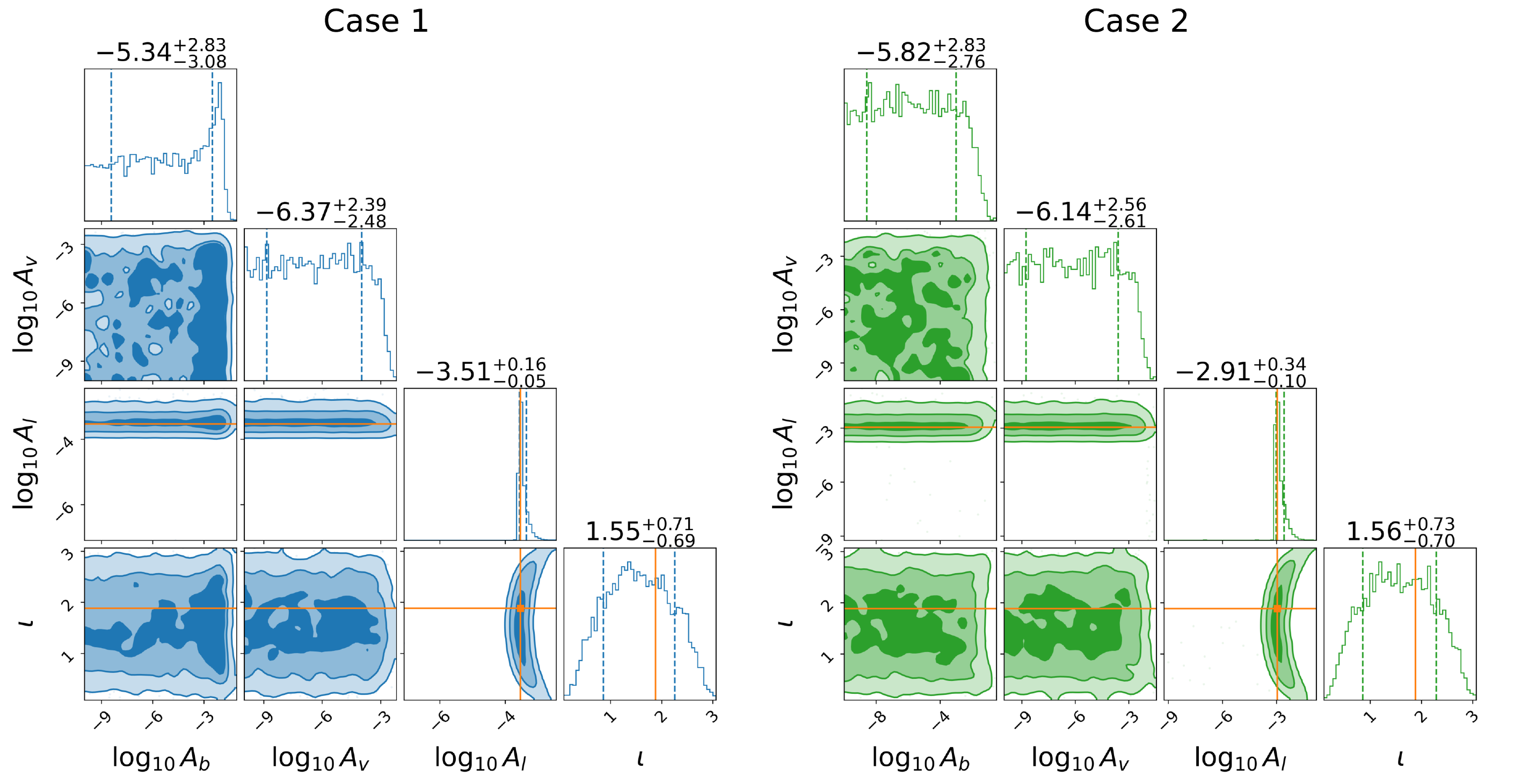}
 \caption{ Similar to Fig.~\ref{fig-bayes-b}. Here, we inject longitudinal mode only.  }
\label{fig-bayes-l}
\end{figure*}

We make use of \textsc{Bilby} \cite{Ashton:2018jfp} with \textsc{dynesty}
\cite{Speagle:2019ivv} as the sampler to do the parameter estimation.  As
pointed by \citet{OBeirne:2019lwp}, there is degeneracy between $\psi$ and
$\Phi_0$ in the parametrization model we use.  Since, in this study, we are more
interested in the relative amplitude and the inclination angle, we only show the
posterior distribution of $\log_{10}A_b$, $\log_{10}A_v$, $\log_{10}A_l$ and
$\iota$ in Figs.~\ref{fig-bayes-b}, \ref{fig-bayes-v} and \ref{fig-bayes-l}, after marginalizing over $\psi$ and $\Phi_0$.
It should be noted that there is strong degeneracy between the relative
amplitude and the inclination angle in the waveform model. Therefore, we get a
V-shape contour of the posterior distribution between $\iota$ and $\log_{10}A_b$
or $\log_{10}A_l$ when we inject only the breathing or longitudinal mode
respectively.  
Thus, we cannot recover the amplitude and the inclination angle separately in both cases.
While for the vector
modes, since the vec-x mode and vec-y mode have different dependences on the
inclination [see Eqs.~\eqref{eq-waveform1} and \eqref{eq-waveform2}], this degeneracy can be broken.

As the response functions of the extra modes are not exactly orthogonal to the response function of the tensor mode, the extra modes will inevitably be weakened after the projection. 
Nonetheless, the parameters of the extra modes still can be recovered to some extent, if the signals are strong enough. 
\citet{OBeirne:2019lwp}  used the original data stream to estimate the parameters after injected both tensor modes and nontensorial modes. The degeneracy between different polarizations leads to larger uncertainty but the results are unbiased. In our study, we used the null stream to estimate the parameters and also got some unbiased results. We think it is an alternative way to estimate the parameters of the extra modes by projecting out the tensor modes. Although there are fewer parameters to recover after the projection,  some part of the information of the extra modes are lost. A more throughout and systematic comparison between the parameter estimation with null stream and that with original stream will be very interesting and valuable in the future.

\section{Discussions}
\label{sec-dis}

In this work, we estimate the detectability of nontensorial polarizations with
PTAs.  We consider a circular SMBBH whose orbit barely evolves during the
observation period.  Two cases for the nontensorial polarizations are
considered, where the dipole radiation dominates in Case 1 and the quadrupole
radiation dominates in Case 2.  Specifically, the waveforms for the two cases
are parameterized in Eqs.~\eqref{eq-waveform1} and \eqref{eq-waveform2}
respectively.  The strength of the extra polarizations is modulated by the
relative amplitude $A_e$.  For the PTA configuration, we consider 49 pulsars
with the same timing variance $\sigma_t=100 \, {\rm ns}$ and we assume a
9.5-year observation with 300 TOAs for each pulsar.  The response of such a PTA
to different polarizations is shown in Fig.~\ref{fig-response}.

We follow \citet{Zhu:2015tua} to construct null streams, in which all the tensor
polarization signals are eliminated.  Then we can quantify the deviation in the
null streams to Gaussian noise to check if there is any evidence of nontensorial
GW signals.  With a frequentist analysis, we first estimate what fraction of the
sky we can detect extra polarizations at $>3\sigma$ confidence level for a
specific strength.  Equivalently, we estimate how likely we can detect
nontensorial polarizations for a randomly distributed source using simulations.
According to the results in Fig.~\ref{fig-location}, the values vary case by
case in the simulations due to the random inclination and noise realization.
Considering a fraction of $50\%$, i.e. when there is a $~50\%$ probability of
finding evidence of nontensorial signals at $>3\sigma$ confidence level, the
median values are $A_b\approx 0.18$, $A_v \approx 0.08$, $A_l\approx0.06$ for
Case 1.  For Case 2, the corresponding median values are $A_b\approx0.96$,
$A_v\approx 0.20$, $A_l\approx 0.22$.  If we include more pulsars, the relative
amplitude can be much lower.

We also project out all of the polarizations, but leave  only one (breathing,
vector, or longitudinal) mode in the null streams.  In this case, we can claim
evidence of a specific nontensorial polarization, that has not been projected
out, if we find significant deviation in the null streams from Gaussian noise.
By setting the injected SNR to be 22, the median values for the relative
amplitude are $A_b= 0.26$, $A_v=0.13$, $A_l=0.15$ for Case 1, and $A_b=1.36$,
$A_v=0.37$, $A_l=0.81$ for Case 2.  The variance of the longitudinal mode is
largest, since its response function strongly depends on the sky locations.

Lastly, we perform Bayesian analysis with the null streams to check how well we
can recover the parameters of the model.  Due to the degeneracy between the
amplitude and the inclination, we can not recover them separately if we only
inject breathing or longitudinal mode, while the combination of the vec-x mode
and vec-y mode can break this degeneracy. 
Although some parts of the extra modes are removed after the projection, we still get unbiased results of parameter estimation with null stream.

In this work, we assume that the SMBBH is in a circular orbit, while for an
eccentric binary, it can emit multiple harmonics of GWs.  Including multiple
harmonics can help break the degeneracy between the amplitude and the
inclination.  Furthermore, it is more realistic to consider the frequency
evolution, so that the Earth term and the pulsar term have different frequency
dependence~\cite{Janssen:2014dka}.  Finally, we model the pulsar data with white
noise only and assume that they are homogeneous. More realistic PTA data may
impact the detectability of nontensorial polarizations.  We leave these
extensive studies to future work.

Finally, in addition to the individual supermassive binary black hole, another important source of continuous waves is isolated neutron stars. 
As an asymmetric body, a rotating neutron star emits GWs in the audio frequency band, which is one of the important targets for ground-based detectors. 
The Bayesian framework to search for extra polarizations from an isolated neutron star has been developed by \citet{Isi:2015cva,Isi:2017equ} and then applied to the real data analysis in Ref.~\cite{LIGOScientific:2017ous}. 
It will be interesting and valuable to apply the null stream method to search for extra polarizations in the continuous waves from an isolated neutron star in the future.

\acknowledgments 
We acknowledge the anonymous referees for their helpful comments to improve our manuscript.
We thank Nataliya Porayko and Lorenzo Speri for providing their codes to simulate
pulsars distributed in the Galactic plane. 
We thank Kejia Lee, Zexin Hu and Ziming Wang for the helpful discussions.
This work was supported by the National Natural Science Foundation of China (12405065, 11991053, 12465013, 11975027, 11721303, 12250410246),   
the National SKA Program of China (2020SKA0120300), 
the Beijing Natural Science Foundation (1242018), 
the China
Postdoctoral Science Foundation (2021TQ0018, 2023M742297),
the Max Planck Partner Group Program funded by the Max Planck Society, 
the Fundamental Research Funds for the Central Universities and the High-performance Computing Platform of Peking University. 


%

\end{document}